\documentclass[twocolumn,showpacs,fleqn,nobibnotes]{revtex4}

\usepackage{amsmath}
\usepackage{graphicx}
\usepackage{float}

\def\lsim{\raise0.3ex\hbox{$<$\kern-0.75em\raise-1.1ex\hbox{$\sim$}}}
\def\gsim{\raise0.3ex\hbox{$>$\kern-0.75em\raise-1.1ex\hbox{$\sim$}}}

\newcommand{\rr}{\mbox{\boldmath $r$}}

\newcommand{\rb}{\mbox{\boldmath $b$}}

\newcommand{\be}{\begin{equation}}
\newcommand{\ee}{\end{equation}}

\begin{document}

\title{A comparative study of the neutrino-nucleon cross section at ultra high energies}
\author{V.P. Gon\c{c}alves
and P. Hepp}

\affiliation{Instituto de F\'{\i}sica e Matem\'atica, Universidade Federal de
Pelotas\\
Caixa Postal 354, CEP 96010-900, Pelotas, RS, Brazil.}

\begin{abstract}
The high energy neutrino cross section is a crucial ingredient in the calculation of the event rate in high energy neutrino telescopes. Currently there are several approaches which predict different behaviours for its magnitude for ultrahigh energies. In this paper we present a comparison between the predictions based on linear DGLAP dynamics, non-linear QCD  and in the imposition of a  Froissart-like behaviour at high energies. In particular, we update the predictions based on the Color Glass Condensate, presenting for the first time the results for  $\sigma_{\nu N}$ using the solution of the running coupling Balitsky-Kovchegov equation. Our results demonstrate that the current theoretical uncertainty for the neutrino-nucleon cross section reaches a factor three for neutrinos energies around  $10^{11}$ GeV and increases to a factor five for $10^{13}$ GeV.  

\end{abstract}

\maketitle

\section{Introduction}

The study of ultra high energy (UHE) cosmic neutrinos provides an opportunity for study particle physics beyond the reach of the LHC \cite{neu_review}. In particular, the Pierre Auger Observatory (PAO)  is sensitive to neutrinos of energy $\ge 10^8$ GeV \cite{pao}.
A crucial ingredient in the calculation of attenuation of neutrinos traversing the Earth and the event rate in high energy neutrino telescopes
is the  high energy neutrino-nucleon cross section, which  provides a probe of Quantum Chromodynamics (QCD) in the kinematic region of very small values of Bjorken-$x$. 
The typical $x$ value probed is $x \approx m_W^2/2m_NE_{\nu}$, which implies that for $E_{\nu} \approx 10^8 - 10^{10}$ GeV one have $x \approx 10^{-4} - 10^{-6}$ at $Q^2 \approx 10^4$ GeV$^2$. This kinematical range was not explored by the HERA measurements of the structure functions \cite{hera}.

The description of QCD dynamics in the high energy limit  still is a subject of intense debate (For a recent review see e.g. Ref. \cite{hdqcd}).
Theoretically, at high energies (small Bjorken-$x$)  one
expects the transition of the regime described by the linear
dynamics, where only the parton emissions are considered, to a new
regime where the physical process of recombination of partons becomes
important in the parton cascade and the evolution is given by a
non-linear evolution equation.  This regime is characterized by the
limitation on the maximum phase-space parton density that can be
reached in the hadron wavefunction (parton saturation), with the
transition being specified  by a typical scale, which is energy
dependent and is called saturation scale $Q_{\mathrm{s}}$  \cite{hdqcd}.
Moreover,  the growth of the parton distribution is expected to saturate, forming a  Color Glass Condensate (CGC), whose evolution with energy is described by an infinite hierarchy of coupled equations for the correlators of  Wilson lines \cite{BAL,CGC}.  
In the mean field approximation, the first equation of this  hierarchy decouples and boils down to a single non-linear integro-differential  equation: the Balitsky-Kovchegov (BK) equation \cite{BAL,KOVCHEGOV}.
Experimentally, possible signals of parton saturation have already
been observed both in  $ep$ deep inelastic scattering at HERA and in deuteron-gold
collisions at RHIC (See, e.g. Ref. \cite{hdqcd,blaizot,universality}).

Existing estimates of the neutrino nucleon cross sections with structure functions constrained by HERA data are based on linear  dynamics  (DGLAP or an unified DGLAP/BFKL evolution) or phenomenological models that mimics the expected behaviour predicted by  the non-linear QCD dynamics. In particular, the neutrino - nucleon cross section was originally calculated at leading order in Ref. \cite{grqs}, with the resulting parametrization being a benchmark for the evaluation of sensitivies of UHE cosmic neutrinos. In Refs. \cite{ancho,ccs} a next-to-leading order analysis was performed, and the uncertainties on high energy $\sigma_{\nu N}$ which are compatible with the conventional DGLAP formalism \cite{dglap} was estimated. Moreover, in Ref. \cite{fjkpp} it was estimated considering an analytical solution of the DGLAP equation, valid at  twist-2 and small-$x$.  These approaches imply a power-like increase of the cross section at ultrahigh energies which is directly associated to the DGLAP solution at small-$x$. In contrast, in Refs. \cite{bbmt,bhm} the HERA data were successfully fitted assuming that the proton structure function saturates the Froissart bound, which implies $F_2^p \propto \ln^2 (1/x)$.
On the other hand, in Refs. \cite{kutak,jamal,magno,armesto} the contribution of non-linear effects  was estimated considering the  leading order solution of the BK equation and distinct phenomenological models based on saturation physics \cite{jamal,magno}. These non-linear approaches were improved recently with the solution of the BK equation including running coupling corrections and the successful description of the HERA data. This fact motivates a revision  of  previous estimates. 
In particular, we present for the first time the predictions for the neutrino-nucleon cross section obtained using as input the  numerical solution of the  
Balitsky-Kovchegov equation \cite{BAL,KOVCHEGOV} including running coupling corrections 
\cite{kovwei1,javier_kov,balnlo}. Moreover, we present the predictions of some recent phenomenological parametrizations based on saturation physics, which provide 
an economical description of a wide range of data with a few parameters.
 Our main motivation is to provide an update on the neutrino-nucleon cross section in the literature and estimate the theoretical uncertainty in its behaviour at ultra high energies.

This paper is organized as follows. In next section (Section \ref{cross}) we present a brief review of the neutrino-nucleon deep inelastic scattering (DIS) cross section, introducing the main formulae. In Section \ref{dinamica} we discuss the QCD dynamics and the models used in the calculations. In Section \ref{resultados} we  compare the predictions of the different models. Finally, in Section \ref{conc} we summarize our main results and conclusions.

\section{The neutrino - nucleon DIS cross section at high energies}
\label{cross}

Deep inelastic neutrino scattering is described in terms of charged current (CC) and neutral current (NC) interactions, which proceed through $W^{\pm}$ and $Z^0$  exchanges, respectively.   
The total cross sections are given by \cite{book}
\begin{eqnarray}
\sigma_{\nu N}^{CC,\,NC} (E_\nu) = \int_{Q^2_{min}}^s dQ^2 \int_{Q^2/s}^{1} dx \frac{1}{x s} 
\frac{\partial^2 \sigma^{CC,\,NC}}{\partial x \partial y}\,\,,
\label{total}
\end{eqnarray}
where $E_{\nu}$ is the neutrino energy, $s = 2 ME_{\nu}$ with $M$ the nucleon mass, $y = Q^2/(xs)$ and $Q^2_{min}$ is the minimum value of $Q^2$ which is introduced in order to stay in the deep inelastic region. In what follows we assume $Q^2_{min} = 1$ GeV$^2$. Moreover, the differential cross section is given by \cite{book}
\begin{eqnarray} 
\frac{\partial^2 \sigma_{\nu N}^{CC,\,NC}}{\partial x \partial y} = \frac{G_F^2 M E_{\nu}}{\pi} \left(\frac{M_i^2}{M_i^2 + Q^2}\right)^2 \nonumber \\
\times \left[\frac{1+(1-y)^2}{2} \, F_2^{CC,\,NC}(x,Q^2) - \frac{y^2}{2}F_L^{CC,\,NC}(x,Q^2) \right. \nonumber\\
\left. + y (1-\frac{y}{2})xF_3^{CC,\,NC}(x,Q^2)\right]\,\,,
\label{difcross}
\end{eqnarray}
where $G_F$ is the Fermi constant and $M_i$ denotes the mass of the charged of neutral gauge boson. 
The calculation of $\sigma_{\nu N}$ involves  integrations over $x$ and $Q^2$. On the one hand,  as the parton  distributions behaves as $x^{-\lambda}$ ($\lambda > 0$) at low $x$, the $x$ integral becomes dominated by the interaction with partons of lower $x$. 
On the other hand, the $Q^2$ integral remains dominated by $Q^2$ values to the order of the electroweak boson mass squared. For $Q^2$ above $M_W^2$ the integrand behaves as $1/Q^4$ and quickly becomes irrelevant. As the neutral current (NC) interactions are subdominant,  we will consider in what follows, for simplicity, only charged current (CC) interactions.

In the QCD improved parton model the structure functions $F_i(x,Q^2)$ are expressed in terms of the parton distributions on the nucleon, which satisfy the DGLAP  \cite{dglap} and/or BFKL \cite{bfkl}  linear dynamics. On the other hand, an efficient way of introducing non-linear effects is the description of the structure functions considering the color dipole approach in which the DIS to low $x$ can be viewed as a result of the interaction of a color $q\bar{q}$ dipole which the gauge boson fluctuates \cite{nik}.  In this approach the $F_2^{CC,\,NC}$ structure function is expressed in terms of the transverse and longitudinal structure functions, $F_2^{CC,\,NC}=F_T^{CC,\,NC} + F_L^{CC,\,NC}$ which are given by 
\begin{eqnarray}
&\,& F_{T,L}^{CC,\,NC}(x,Q^2)  =  \frac{Q^2}{4\pi^2} \nonumber \\ &\times & \int_0^1 dz \int d^2  \rr |\Psi^{W,Z}_{T,L}(\rr,z,Q^2)|^2 \sigma^{dp}(\rr,x)\,\,
\label{funcs}
\end{eqnarray} 
where $r$ denotes the transverse size of the dipole, $z$ is the longitudinal momentum fraction carried by a quark and  $\Psi^{W,Z}_{T,L}$ are proportional to the wave functions of the virtual charged of neutral gauge bosons corresponding to their transverse or longitudinal polarizations. Explicit expressions for $\Psi^{W,Z}_{T,L}$ are given, e.g., in Ref. \cite{kutak}.    Furthermore, $\sigma^{dp}$ describes the interaction of the  color dipole with the target. In next section we will discuss some models for $\sigma^{dp}$, based on the non-linear QCD dynamics, which describe the current HERA data.

A comment is in order. The HERA measurements of the structure function  at low - $x$ ($x  \approx 10^{-6}$) are for very low values of $Q^2$ ($Q^2 \ll 1$ GeV$^2$), which implies that  small $x$ extrapolations of the parton distributions are necessary to estimate $\sigma_{\nu N}$ above $E_{\nu} \approx 10^7$ GeV. As we will discuss in more detail in the next sections, these extrapolations contain significant uncertainties.

\section{High Energy QCD Dynamics}
\label{dinamica}

High energy neutrino-nucleon cross section accesses very large values of $Q^2$ and very small values of Bjorken $x$. 
 In the last years, it has been calculated considering different theoretical approaches with structure functions constrained by HERA data. 
It was originally calculated using the LO DGLAP equation in Ref. \cite{grqs} (GQRS model).
In Ref. \cite{ccs} (C-SS model) a next-to-leading order analysis was performed considering a  global fit of the ZEUS data and the uncertainties on high energy $\sigma_{\nu N}$ which are compatible with the conventional DGLAP formalism  was estimated. 
Another approach based on DGLAP dynamics was proposed in \cite{fjkpp} (FJKPPP model), where $\sigma_{\nu N}$ was estimated using an analytic result for the DGLAP evolution of the structure functions, valid at twist-2 in the region of small-$x$ and for   a soft non-perturbative input.  As the DGLAP equation can break down at low $x$ because of the $\ln (1/x)$ terms which appear in the perturbative series, an alternative approach which incorporates both the $\ln (1/x)$ resummation and the complete LO DGLAP evolution was proposed in \cite{kma}, which we denote unified DGLAP-BFKL model hereafter (See also Ref. \cite{pena}).

These predictions are based on the linear DGLAP and/or BFKL equations which implies a power increase with the energy  of the neutrino-nucleon cross section that eventually violate the Froissart bound.
In Ref. \cite{bbmt} the UHE $\sigma_{\nu N}$ was estimated using a phenomenological approach (BBMT model) which considers the imposition of the Froissart unitarity and analyticity constraints on inclusive deep-inelastic cross sections. It implies  that  $F_2 \propto \ln^2(1/x)$  at very small $x$. In \cite{bbmt} very good fits to data were obtained for $x< 0.1$ and a wide range of $Q^2$.
More recently, in \cite{bhm},  this model was updated considering the most recent analysis of the complete ZEUS and H1 datasets from HERA. In what follows we denote by BHM the corresponding predictions.

Another alternative to estimate the ultrahigh energy behavior of the neutrino-nucleon cross section is to express the structure functions in the dipole approach [Eq. (\ref{funcs})] and to consider the state-of-art of the non-linear QCD dynamics: the Color Glass Condensate formalism. In this formalism, the dipole - target cross section 
$\sigma^{dp}$ can be computed in the eikonal approximation,
resulting
\begin{eqnarray}
\sigma^{dp} (x,\rr)=2 \int d^2 \rb \, {\cal{N}}(x,\rr,\rb)\,\,,
\end{eqnarray}
where ${\cal{N}}$ is the  dipole-target forward scattering amplitude
for a given impact parameter $\rb$  which encodes all the
information about the hadronic scattering, and thus about the
non-linear and quantum effects in the hadron wave function. It is
useful to assume that the impact parameter dependence of $\cal{N}$
can be factorized as ${\cal{N}}(x,\rr,\rb) = {\cal{N}}(x,\rr)
S(\rb)$, so that $\sigma^{dp}(x,\rr) = {\sigma_0}
\,{\cal{N}}(x,\rr)$, with $\sigma_0$ being   a free parameter
related to the non-perturbative QCD physics. The Balitsky-JIMWLK
hierarchy  describes the energy evolution of the dipole-target
scattering amplitude ${\cal{N}}(x,\rr)$.
In the mean field approximation, the first equation of this  hierarchy decouples and boils down to the Balitsky-Kovchegov (BK) equation \cite{BAL,KOVCHEGOV}.

In the last years 
the next-to-leading order corrections to the  BK equation were calculated  
\cite{kovwei1,javier_kov,balnlo} through the ressumation of $\alpha_s N_f$ contributions to 
all orders, where $N_f$ is the number of flavors. Such calculation allows one to estimate 
the soft gluon emission and running coupling corrections to the evolution kernel.
The authors have verified that  the dominant contributions come from the running 
coupling corrections, which allow us to  determine the scale of the running coupling in the 
kernel. The solution of the improved BK equation was studied in detail in Refs. 
\cite{javier_kov,javier_prl}. Basically, one has that the running of the coupling reduces 
the speed of the evolution to values compatible with experimental data, with the geometric 
scaling regime being reached only at ultra-high energies. In \cite{bkrunning} a global 
analysis of the small $x$ data for the proton structure function using the improved BK 
equation was performed  (See also Ref. \cite{weigert}). In contrast to the  BK  equation 
at leading logarithmic $\alpha_s \ln (1/x)$ approximation, which  fails to describe the HERA 
data, the inclusion of running coupling effects in the evolution renders the BK equation 
compatible with them (See also \cite{vic_joao,alba_marquet,vicmagane}).

\begin{figure}[t]
\includegraphics[scale=0.35]{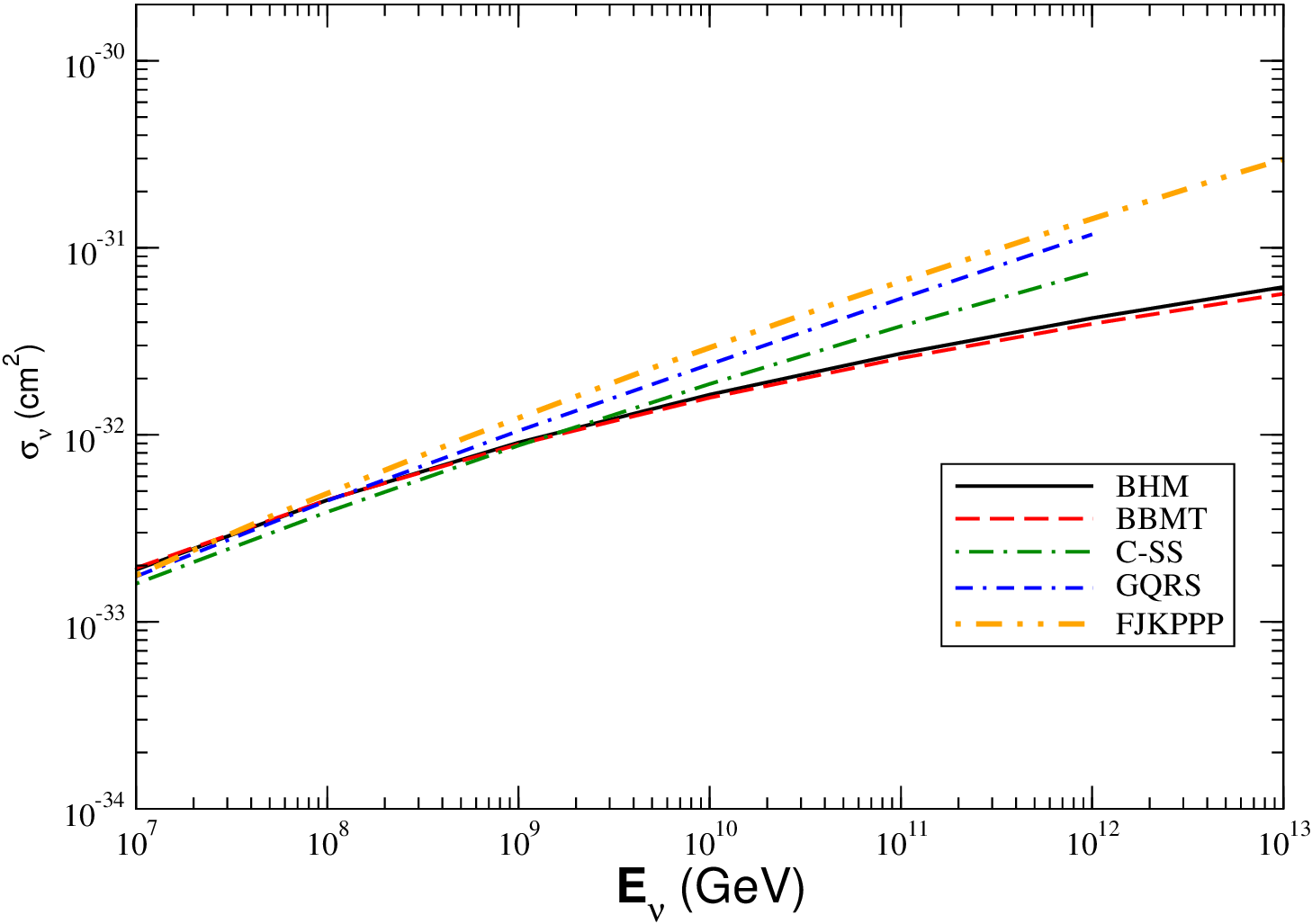}
\caption{(Color online) The neutrino nucleon CC cross section as a function of the neutrino energy, $E_{\nu}$. Comparison between the linear and Froissart-inspired approaches.}
\label{fig:1}
\end{figure}

The dipole-target cross section  can also be calculated considering  phenomenological parametrizations for ${\cal{N}}(x,r)$ based on saturation physics, which provide 
an economical description of a wide range of data with a few parameters.
Several models for the forward dipole cross section have been used in the literature 
in order to fit the HERA and RHIC data \cite{dipolos,dipolos2,dipolos3,dipolos4,dipolos5,dipolos6,dipolos7,soyez,iim,kkt,dhj,universality,buw}. In general, the   dipole scattering amplitude is modelled in the coordinate space in terms of a simple Glauber-like formula as follows
\begin{eqnarray}
{\cal{N}}(x,r) = 1 - \exp\left[ -\frac{1}{4} (r^2 Q_s^2)^{\gamma (x,r^2)} \right] \,\,,
\label{ngeral}
\end{eqnarray}
where   $\gamma$ is the anomalous dimension of the target gluon distribution.    The main difference among the distinct phenomenological models comes from the  predicted behaviour for the anomalous dimension, which determines  the  transition from the non-linear to the extended geometric scaling regimes, as well as from the extended geometric scaling to the DGLAP regime (See e.g. \cite{hdqcd}).  The current models in the literature consider the general form $\gamma = \gamma_s + \Delta \gamma$, where $\gamma_s$ is the anomalous dimension at the saturation scale and $\Delta \gamma$ mimics the onset of the geometric scaling region and DGLAP regime. One of the basic differences between these models is associated to the behaviour predicted for $\Delta \gamma$. While the models proposed in Refs. \cite{iim,kkt,dhj} assume that $\Delta \gamma$ depends on terms which violate the geometric scaling, i.e. depends separately on $r$ and rapidity $Y = \ln(1/x)$,  the model  proposed in Ref. \cite{buw} (BUW model) consider that it is a function of $r Q_s$.  In particular, these authors  demonstrated that the RHIC data for hadron production in $dAu$ collisions for all rapidities are compatible with geometric scaling and that 
 geometric scaling violations  are not observed at RHIC \cite{buw}. In contrast, the IIM analysis  \cite{iim}  implies that a substantial amount of geometric scaling violations is needed in order to accurately describe the $ep$ HERA experimental data.  
In our analysis we will estimate $\sigma_{\nu N}$ considering the rcBK solution and the IIM and BUW models. Moreover, we will consider  the  improved version of the IIM model proposed in Ref. \cite{soyez} (denoted hereafter IIM-S model), which includes the heavy quark effects on the saturation in the fit of the HERA data. The main differences of the IIM-S model in comparison to the IIM one are the larger value of the anomalous dimension and the  smaller value of the exponent which determines the energy growth of the saturation scale. Moreover, the IIM-S model considers  a newer H1 and ZEUS datasets in the fit, in contrast to the IIM one which only considers the ZEUS data.



\begin{figure}[t]
\includegraphics[scale=0.35]{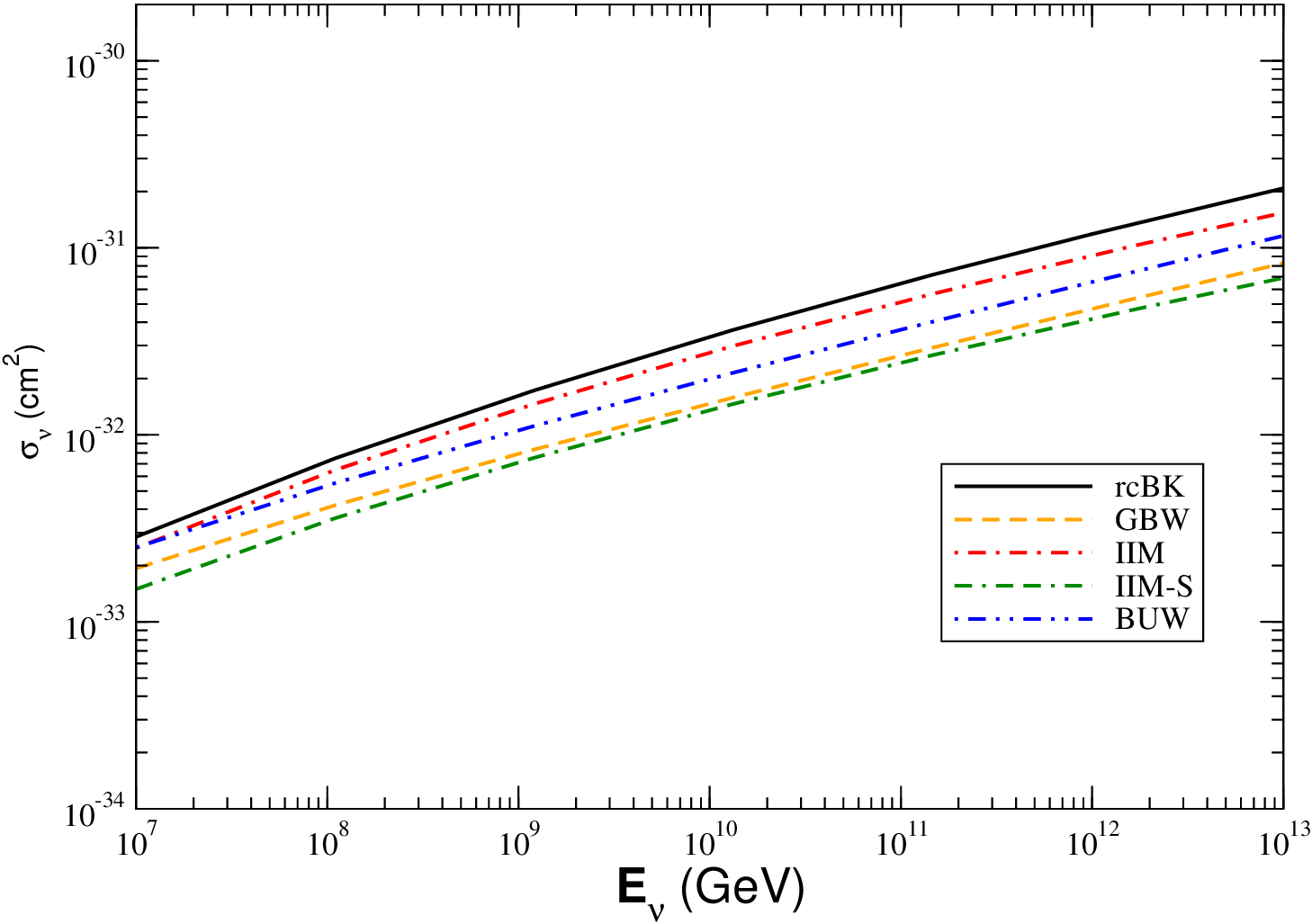}
\caption{(Color online) The neutrino nucleon CC cross section as a function of the neutrino energy, $E_{\nu}$. Comparison between the non-linear approaches.}
\label{fig:2}
\end{figure}

\section{Results}
\label{resultados}

In this section we present a comparison between the predictions of the linear approaches (GQRS, C-SS and FJKPPP), the Froissart-inspired models (BBMT and BHM) and non-linear approaches (rcBK, IIM, IIM-S and BUW). In  Fig. \ref{fig:1} the  energy dependence of the neutrino nucleon CC cross section predicted by the  linear and Froissart-inspired approaches are compared.  As expected from the solution of the DGLAP equation at small-$x$, the GQRS, C-SS and FJKPPP models predict a strong increase of the cross section at ultrahigh energies. Although these approaches agree at low energies, where the behavior of the parton distributions are constrained by the HERA data, they differ by a factor 1.25 at $E_{\nu} = 10^{12}$ GeV. We have that the C-SS prediction, which comes from a global fit of the ZEUS data, can be considered as a lower bound for the linear predictions. On the other hand, the FJKPPP prediction, which considers an analytical solution of the DGLAP equation at small-$x$, implies a stronger increase with the energy  similar to the   GQRS one, largely used  to estimate the event rates in neutrinos telescopes. It is important to emphasize that GQRS parameterization was obtained using a restrict set of experimental data and DGLAP evolution at leading order. In contrast, the BBMT and BHM approaches, which assume a Froissart-like behavior for the structure functions at small-$x$ [$F_2 \propto \ln^2(1/x)$], predict at ultra high energies a cross section smaller than the FJKPPP one by a factor $\approx 3$. We have that the BBMT and BHM predictions, which comes from a fit of the structure functions using different datasets, are very similar.

\begin{figure}[t]
\includegraphics[scale=0.35]{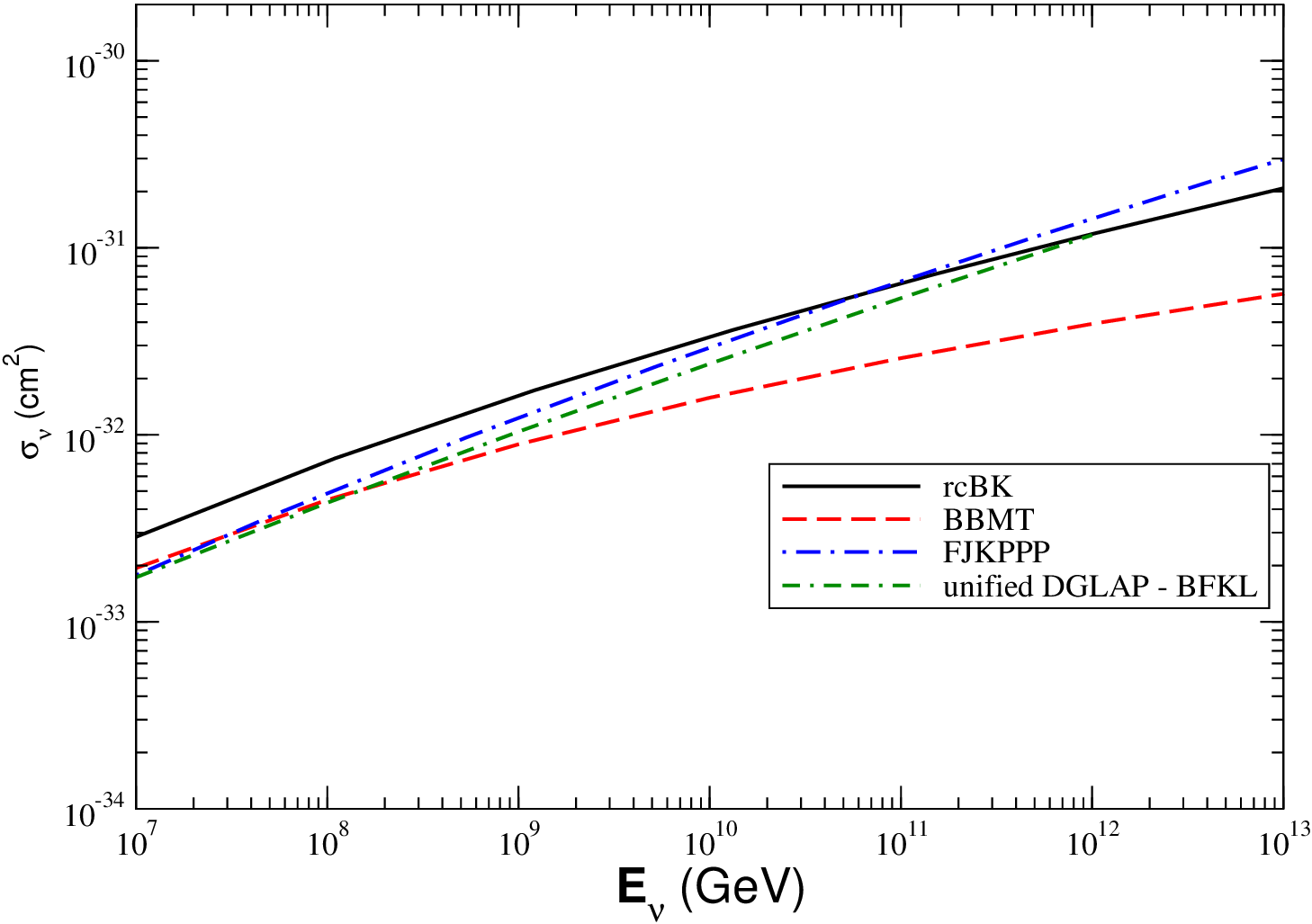}
\caption{(Color online) The neutrino nucleon CC cross section as a function of the neutrino energy, $E_{\nu}$. Comparison between the linear and non-linear approaches.}
\label{fig:3}
\end{figure}

In Fig. \ref{fig:2} we present the predictions of the non-linear approaches for the energy dependence of the neutrino nucleon CC cross section. For comparison, we also includes the GBW model \cite{dipolos} which was used in previous studies of the saturation effects \cite{kutak,magno}.
 We have that the IIM-S prediction is very similar to the GBW one, being a lower bound for the non-linear predictions at ultrahigh energies. In contrast, the rcBK prediction can be considered an upper bound. These predictions differ by a factor $\approx 3$ at $E_{\nu} = 10^{12}$ GeV. The large difference between the IIM and IIM-S predictions is directly associated to the distinct energy dependence for the saturation scale and the treatment of heavy quarks proposed by these  models. It implies a different normalization for the dipole-target cross section and, consequently, for the neutrino-nucleon CC cross section. The IIM, BUW and rcBK predictions are similar at low energies, but differ by a factor 1.8 at ultra high energies.  It is important to emphasize that BUW and rcBK models successfully describe the current RHIC and HERA data \cite{buw,bkrunning,alba_marquet}.
 
Finally, in Fig. \ref{fig:3} we present a comparison between the predictions of linear and non-linear approaches. For comparison we also include the prediction obtained by the unified DGLAP-BFKL approach \cite{kma}, which is similar to the FJKPPP one. We have that the FJKPPP and rcBK predictions are similar for $E_{\nu} = 10^{11}$ GeV and differ by $\approx 15\%$ at $E_{\nu} = 10^{12}$ GeV. In contrast, the rcBK and BBMT differ by a factor $\approx 3$ at $E_{\nu} = 10^{11}$ GeV. The theoretical uncertainty increases for a factor $\approx 5.5$ when we compare the FJKPPP and BBMT predictions for $E_{\nu} = 10^{13}$ GeV. Our results demonstrate that the determination of $\sigma_{\nu N}$ can be useful to contrain the underlying QCD dynamics. In principle, this cross section could be  constrained at high energies by studying the  ratio between quasi-horizontal deeply penetrating air showers and Earth-skimming tau showers \cite{ancho}.

\section{Summary}
\label{conc}
Detection of UHE neutrinos may shed light on the observation of air showers events with energies in excess of $10^{11}$ GeV, reveal aspects of new physics as well as of the QCD dynamics at high energies. One of the main ingredients for estimating event rates in neutrino telescopes (e.g. ICECUBE) and  cosmic ray observatories (e.g. AUGER) 
is the neutrino - nucleon cross section. 
In this paper we examined to what extent the cross section is sensitive to the presence of new dynamical effects in the QCD evolution. 
We compare the predictions of several approaches based on different assumptions for the QCD dynamics. In particular, we have compared the more recent predictions based on the  NLO DGLAP evolution equation with those from the CGC physics obtained using the running coupling BK solution or phenomenological models. Our results demonstrate that the current theoretical uncertainty for the neutrino-nucleon cross section reaches a factor three for neutrino energies around  $10^{11}$ GeV and increases to 5.5 for   $E_{\nu} = 10^{13}$ GeV.

A final comment is in order. In this paper we estimated the range of possible values for $\sigma_{\nu N}$ at large energies within the Standard Model. It makes possible to search for enhancements in the neutrino-nucleon cross section due to physics beyond the perturbative SM (See, e.g. \cite{ancho2}).




\begin{acknowledgments}
This work was  partially financed by the Brazilian funding
agencies CNPq, CAPES and FAPERGS.
\end{acknowledgments}

\end{document}